\documentclass[floatfix,aps,prd,showpacs,amsmath,amssymb,nofootinbib,eqsecnum,superscriptaddress]{revtex4}

\usepackage{epsfig}

\begin{document}

\title{Vortex Condensation in the Dual Chern-Simons Higgs Model}

\author{Rudnei O. Ramos} \email{rudnei@uerj.br} \affiliation{Departamento de
  F\'\i sica Te\'orica, Universidade do Estado do Rio de Janeiro, 20550-013,
  Rio de Janeiro, RJ, Brazil}

\author{J. F. Medeiros Neto} \email{jfmn@ufpa.br} \affiliation{Faculdade de
  F\'\i sica, Universidade Federal do Para, 66075-110, Bel\'em, PA, Brazil}

\begin{abstract}
  
  The contribution of nontrivial vacuum (topological) excitations, more
  specifically vortex configurations of the self-dual Chern-Simons-Higgs
  model, to the functional partition function is considered. By using a
  duality transformation, we arrive at a representation of the partition
  function in terms of which explicit vortex degrees of freedom are coupled to
  a dual gauge field. By matching the obtained action to a field theory for
  the vortices, the physical properties of the model in the presence of vortex
  excitations are then studied. In terms of this field theory for vortices in
  the self-dual Chern-Simons Higgs model, we determine the location of the
  critical value for the Chern-Simons parameter below which vortex
  condensation can happen in the system.  The effects of self-energy quantum
  corrections to the vortex field are also considered.

\end{abstract}


\maketitle

\section{Introduction}

Gauge field theories in two spatial dimensions have long been recognized as
important for understanding several physical phenomena that can be well
approximated as planar ones, like high temperature superconductors and the
fractional quantum Hall effect.  In particular, Chern-Simons (CS) gauge
theories have a special place in understanding these phenomena, not to mention
the interest in their theoretical aspects on their own (for a review, see e.g.
Ref.~\cite{review} and references therein).  CS gauge theories exhibit a
number of interesting properties. {}For example, it provides a mass term for
the gauge field, while keeping renormalizability, without evoking spontaneous
symmetry breaking. It can have the effect of statistical transmutation,
attaching magnetic fluxes to a fermion or boson coupled to the gauge field and
making them anyons~\cite{anyons}.

Another important aspect regarding CS gauge theories, when coupled to symmetry
broken scalar potentials, is the existence of both topological and
nontopological vortex solutions~\cite{khare}.  These vortices are charged and
anyon-like solutions that may be of relevance in explaining several phenomena
in planar condensed matter systems, like in high temperature superconductors
and the fractional quantum Hall effect, already mentioned above.  Vortex
solutions have been shown to exist in both a Maxwell-Chern-Simons-Higgs (MCSH)
model (where both a kinetic Maxwell term and a CS term are present) as well in
the Chern-Simons-Higgs (CSH) model (in the absence of a Maxwell term), where
the vortices can have the property of self-duality, with the field equations
reducing to first order differential equations~\cite{pac,jackiw}.

As far the physics of topological excitations like vortices are concerned, one
important question is related to the possibility of condensation of these
excitations in the system under appropriate conditions. Under these
circumstances most of the physical properties of the system should be
determined by those of the condensate.  Condensation of topological
excitations are believed to have relevance in the interpretation of many
physical phenomena, like for instance in the confinement picture in dual
formulations of gauge field theories~\cite{chernodub}.  We should also recall
that there are many examples of physical systems in which phase transitions
can be driven by topological excitations in quantum field theory as well as in
condensed-matter physics~\cite{kleinertI-II}, which makes the study of
condensation of nontrivial vacuum excitations of relevance in different
contexts.

It is known that the CSH model exhibits a phase transition between a vortex
condensed phase and one in which the vortices are not
condensed~\cite{olesen,caffarelli}.  In this work we consider the vortex
condensation in the CSH model specialized to the case of the self-dual
potential for the scalar field~\cite{pac,jackiw}, in which case vortices can
be considered as noninteracting.  This is used only for convenience, since
then explicit expressions for the vortex energy and the dynamical mass for the
vortices follow.  Other generic potentials could well be used, provided that
it does not depend on the phase of the scalar field $\phi$ (so depends only on
the product $\phi\phi^*$) and has a minimum at a non-zero value of $\phi$.
However, exact relations are not known for these more general potentials like
for the self-dual one.  Vortex condensation in CS theories has been considered
before, in the context of the self-dual models, in
Refs.~\cite{olesen,caffarelli}.  In particular, in Ref.~\cite{caffarelli}, it
was formally shown that vortices should condense below some critical value for
the CS coupling parameter.  However, no exact prediction for this value of the
CS parameter was made. Also, being this condensation a transition point in the
system, a determination of the order of this transition is also still lacking.
These are important points that need to be addressed and that we will attempt
to study in this paper, at least in part.

The existence of nontrivial solutions of the vortex type implies that, when
performing the path integral over the fields in the partition function, there
will be contributions coming from field configurations corresponding to these
vortex solutions and, under appropriate conditions (depending on the values of
the model parameters, for example), these solutions can dominate the partition
function over constant vacuum configurations. This would then signal the
possibility of condensation of these nontrivial vacuum excitations.  Our
strategy to study the vortex condensation problem in the CSH model is as
follows. Vortex excitations are made explicit in the functional action by
making use of a series of dual transformations for the original Lagrangian
fields, obtaining an equivalent action, in which it becomes clear the vortex
contributions.  By properly matching our dual action to a field theory model,
it is then possible to write it in terms of a vortex field $\psi$ coupled to a
vectorial field. In this process of writing the action in terms of an explicit
vortex field, a dynamical mass for the vortices is generated. {}From the
dynamical generated mass for the vortices we can then infer, already at the
tree-level, about the possibility of vortex condensation. In this case, there
is a critical value for the CS parameter where the dynamical mass for the
vortices vanishes and these become energetically favorable to condense.  We
also calculate the one-loop self-energy corrections for the introduced vortex
field to check the stability of the condensation point when quantum
corrections are included.

This paper is organized as follows.  In Sec. II we introduce the CSH model and
briefly review its usual vortex solutions.  In Sec.  III we derive the dual
action, equivalent to the original action, but where the vortex excitation
degrees of freedom can be made explicit. By matching this dual action to a
field theory model, an appropriate vortex field can then be introduced.
Working in the London approximation, where scalar field fluctuations are
frozen, we obtain an explicit expression for the dynamical mass term for the
vortex field that emerges naturally in the passage from the vortex coordinates
to the vortex field.  The point for vortex condensation is then derived from
the dynamical mass term for the vortex field.  In Sec. IV we compute the
self-energy corrections for the vortex field in the dual model and investigate
the change of the tree-level vortex condensation point when one-loop quantum
corrections are considered.  Our conclusions and final remarks are given in
Sec. V.

\section{The Chern-Simons-Higgs model and its Vortex Solution}
\label{model}

Let us consider the Abelian CSH model, defined in terms of a complex scalar
field $\phi$ and Abelian gauge field $A_\mu$. The quantum partition function
can be written as (throughout this work we work in the Euclidean space-time)

\begin{equation}
Z = \int \mathcal{D}A_{\mu}\mathcal{D}\phi \mathcal{D}\phi^* \exp
\left\{ - S_E[A_\mu, \phi,\phi^*] \right\}\;,
\label{ZAphi}
\end{equation}
with Euclidean action given by (indices run from 1 to 3)

\begin{equation}
S_E[A_\mu, \phi,\phi^*] = \int d^3 x \left[ - i\frac{\theta}{4}
{}\epsilon_{\mu \nu \gamma}A_{\mu}{}F_{\nu \gamma} + |D_\mu
\phi|^2 + V(|\phi|) \right] \;,
\label{actionE}
\end{equation}
with $D_\mu \equiv \partial_\mu + ieA_\mu$ and $\theta$ is the CS parameter.
$V(|\phi|)$ is a symmetry breaking polynomial potential, independent of the
phase of the complex scalar field.  The potential $V(|\phi|)$ is some
potential with a nonvanishing symmetry breaking minimum for the scalar field
$\phi$, $|\langle \phi \rangle| = v \neq 0$.  {}For example, we can have the
usual quartic symmetry-broken scalar potential,

\begin{equation}
V(|\phi|) =  \frac{\lambda}{4}\left( |\phi|^2 - v^2\right)^2\;,
\label{V1}
\end{equation}
or the sixth-order self-dual potential \cite{jackiw},

\begin{equation}
V(|\phi|) =  \frac{e^4}{\theta^2}\left( |\phi|^2 -
v^2\right)^2|\phi|^2\;,
\label{V2}
\end{equation}
with degenerate minima at $|\phi|^2 =v^2$ and $|\phi| =0$.  Note also that in
$2+1$ dimensions, a sixth-order potential in the scalar field still gives a
renormalizable theory. In the following we will consider the potential
(\ref{V2}) that leads to the so-called dual vortex solutions~\cite{jackiw}.

The field equations corresponding to Eq. (\ref{actionE}) are known to have
finite energy field solutions corresponding to vortices, that carry an
electric charge $Q$, given in terms of the magnetic flux $\Phi$ by

\begin{equation}
Q = \theta \int d^2x F_{12} \equiv \theta \Phi\;,
\label{Q}
\end{equation}
and that they are also anyons, with a spin $j= Q \Phi/(4 \pi)$ \cite{anyons}.
The field equations, though not having exact solutions, admit radially
symmetric like solutions for a vortex, with multiplicity $n$, given by (using
polar coordinates denoted here by $r,\chi$)

\begin{eqnarray}
\phi_{\rm vortex} &=& \varphi(r) \,\exp(i n \chi)\;,
\label{phi vortex}
\\
A_{\mu, {\rm vortex}}  &=& \frac{n}{e} A(r)\; \partial_\mu \chi\;,
\label{A vortex}
\end{eqnarray}

\noindent
where the functions $\varphi(r)$ and $A(r)$ vanish at the origin and have the
asymptotic behavior $\varphi(r \to \infty) \to v$ and $A(r \to \infty) \to 1$.
The functions $\varphi(r)$ and $A(r)$ are obtained (numerically) by solving
the classical field equations. Then, from Eqs. (\ref{phi vortex}) and (\ref{A
  vortex}), we see that at spatial infinity the scalar field $\phi$ goes to
the vacuum $v$, while the gauge $A_\mu$ becomes a pure gauge.  In this case,
the flux $\Phi$ in Eq. (\ref{Q}) becomes $\Phi =2 \pi n/e$. Also, since the
scalar field must be a single-valued quantity, Eq. (\ref{Q}) implies that, on
the vortex, the phase $\chi$ must be singular.  Therefore, the phase $\chi$
can be separated into two parts: in a regular part, $\chi_{\rm reg}$, and in a
singular one, $\chi_{\rm sing}$, due to the vortex configuration.

The energy of the n-vortex excitations is determined by using the solutions
(\ref{phi vortex}) and (\ref{A vortex}) in the static energy action functional
that is given by

\begin{equation}
E=\int d^2 x \left[ \left| D_i \phi\right|^2 -e^2 A_0^2 |\phi|^2 -
\theta A_0 F_{12} + V(|\phi|) \right]\;,
\label{E}
\end{equation}
which, upon the use of the vortex solutions, gives \cite{jackiw}

\begin{equation}
E_{\rm vortex} \geq 2\pi  v^2 |n| \;,
\label{Evort}
\end{equation}
with equality fulfilled when the self-duality equations for the gauge field
and the scalar field with the self-dual potential (\ref{V2}), are satisfied.
Equation (\ref{Evort}) shows in particular that self-dual vortices are
noninteracting. Special cases of interacting vortex solutions can be generated
by modifications of the self-dual potential (\ref{V2})~\cite{khare2}.

\section{The Dual-transformed action}
\label{dual}

We now describe the steps necessary to make explicit in the model action Eq.
(\ref{actionE}) the vortex degrees of freedom. We start by writing the complex
scalar field in a polar-like parametrization form as $\phi = \rho \exp(i
\chi)/\sqrt{2}$.  {}From the discussion in the previous section, this implies
that on the vortices, the phase $\chi$ is a multivalued function and $\chi$ in
general can be expressed in terms of a regular (single valued) and a singular
part as $\chi(x) = \chi_{\rm reg}(x) + \chi_{\rm sing}(x)$. The quantity

\begin{equation}
{\cal J}^\mu = \frac{1}{2 \pi} \epsilon^{\mu \nu \gamma}
\partial_\nu \partial_{\gamma} \chi_{\rm sing}\;,
\label{vJ}
\end{equation}
defines the vortex current~\cite{lee}.

The existence of the topological vortex solutions imply that, when performing
the path integral over the fields in (\ref{ZAphi}), there will be field
configurations corresponding to these vortex solutions. We next make these
solutions explicit in the functional action.  {}From the modulus and phase
parametrization for the complex scalar field, the partition function
(\ref{ZAphi}) becomes

\begin{eqnarray}
Z&=&\int \mathcal{D}A_\mu \mathcal{D}\rho \mathcal{D}\chi \left(
\prod_x\rho \right) \exp\left\{ - \int d^3x\left[ \frac{1}{2}
\rho^2\left( \partial_\mu \chi +e A_\mu \right)^2 -
i\frac{\theta}{2} {}\epsilon_{\mu \nu \gamma}A_{\mu}{}\partial_\nu A_\gamma
+\frac{1}{2}\left( \partial_\mu \rho \right)^2 + V(\rho) \right]
\right\} \;.
\label{ZArhochi}
\end{eqnarray}
The functional integration over the regular phase $\chi_{\rm reg}(x)$
dependent terms in the partition function is performed as follows,

\begin{eqnarray}
\lefteqn{\int \mathcal{D}\chi \,\exp \left[ -\int d^3 x
\frac{1}{2} \rho^2\left(
\partial_\mu \chi +e A_\mu \right)^2\right] }  \nonumber \\
&=&\int \mathcal{D}\chi_{\mathrm{sing}}\,\mathcal{D}\chi_{\mathrm{reg}}
\mathcal{D}C_\mu \left( \prod_x\rho^{-3}\right) \,\exp \left\{ -\int
d^3 x\left[ \frac{1}{2\rho^2}C_\mu ^2-iC_\mu \left( \partial_\mu
\chi _{\mathrm{reg}}\right) -iC_\mu \left( \partial_\mu \chi_{\mathrm{sing}
}+eA_\mu \right) \right] \right\}  \nonumber \\
&=&\int \mathcal{D}\chi_{\mathrm{sing}}\left( \prod_x\rho^{-3}\right)
\mathcal{D}h_{\mu}\,\exp \left\{ -\int d^3x\left[ \frac{\kappa}{
16 \pi^2 \rho^2}H_{\mu\nu}^2 -i \frac{e\kappa^{1/2}} {4 \pi}
\epsilon_{\mu \nu \gamma} h_\mu F_{\nu \gamma} -i \kappa^{1/2}
{\cal J}_\mu h_\mu   \right] \right\} \;,
\label{dual2}
\end{eqnarray}

\noindent
where the functional integral over the field $C_\mu$, in the second line, was
introduced in order to linearize the exponent term in the first line.  The
functional integral over $\chi_{\mathrm{reg}}$ in the second line of Eq.
(\ref{dual2}) can now be easily done.  It gives a constraint on the functional
integral measure, $\delta (\partial _\mu C_\mu )$, which can be represented in
a unique way by expressing the field $C_\mu $ in terms of a dual field, $C_\mu
= \frac{\kappa^{1/2}}{2\pi} \epsilon_{\mu \nu \gamma} \partial_\nu h_{\gamma}
\equiv \frac{\kappa^{1/2}}{4 \pi} \epsilon_{\mu \nu \gamma}H_{\nu \gamma}$,
where $\kappa $ is some arbitrary parameter with mass dimension and $H_{\mu
  \nu}= \partial_\mu h_\nu - \partial_\nu h_\mu$.

The functional integral over the original gauge $A_\mu$ in Eq.
(\ref{ZArhochi}) can also be immediately performed by using Eq. (\ref{dual2}),
from which we then obtain for Eq. (\ref{ZArhochi}) the result

\begin{eqnarray}
Z&=&  \int D \chi_{\rm sing} D \rho \, (\prod_x \rho^{-2}) D h_\mu
\exp(-  S)\;,
\label{finalZ}
\end{eqnarray}
where

\begin{eqnarray}
S&=& \int d^3 x  \left[ \frac{m^2}{16 \pi^2 e^2
\rho^2}H_{\mu\nu}^2 -i \frac{m}{e} {\cal J}_\mu h_\mu +  i
\frac{m^2}{8 \pi^2 \theta} \epsilon_{\mu \nu \gamma} h_\mu
\partial_\nu h_\gamma
\right.   \nonumber \\
&+&  \left.  \frac{1}{2}\left( \partial_\mu \rho \right)^2 +
V(\rho)      \right]  \;.
\label{S}
\end{eqnarray}
In Eq. (\ref{finalZ}) an overall field independent multiplicative factor was
omitted and in Eq. (\ref{S}) we have defined for convenience a new parameter
$m = e \kappa^{1/2}$, with mass dimension. This arbitrary mass parameter $m$
in the final dual action is just a spurious constant that can be absorbed in a
redefinition of the dual gauge field $h_\mu$ and none of our results will
depend on it.  Note also that the gauge invariance of the original model,
$\delta A_\mu = \partial_\mu \Lambda(x)$, is now replaced in the dual action
in Eq.  (\ref{S}) by $\delta h_\mu(x) =\partial_\mu \Lambda (x)$. Thus, gauge
invariance, now in terms of the $h_\mu$ field, is still preserved.

The obtained expression for the partition function of the dual CSH model makes
evident the contribution due to vortices in the path integral. In comparing
Eqs. (\ref{finalZ}) and (\ref{S}) with the result obtained by the authors of
Ref.  \cite{lee} for the CSH model, we note that their dual result, in the
absence of external fields and currents included in there, is exactly
equivalent to our expression.  Another result that we observe in (\ref{S}) is
the characteristic dualization of the CS coefficient, where from the original
action in (\ref{ZArhochi}) to the dual one Eq. (\ref{finalZ}), it is changed
like $\theta \to - 1/ (4 \pi^2\theta)$.  This dualization for the CS parameter
has also been shown previously, like in Refs. \cite{wen} and \cite{dolan} (in
this last reference it was also shown a detailed derivation of the field
functional derivations leading to this result).  As explained in \cite{lee},
this sign difference of the CS coefficient between the original action and the
dual one is of relevance in interpreting the statistics of the vortices in the
theory as anyons.

{}From Eq. (\ref{S}) we see that the vortex degrees of freedom, represented by
the vortex current (\ref{vJ}), appear coupled with the new gauge field
$h_\mu$. This is a non-vanishing quantity due to the singular nature of
$\chi_{\mathrm{sing}}$ of the Higgs field phase and, hence, this interaction
term will contribute in the action, along with the worldline $x^\mu (\tau)$
swept by the vortex (if taken as pointlike particles).  This can be made more
explicit once we write the vortex current, for unity winding number vortex
excitations ($n=1$), corresponding to the energetically dominant
configurations, in the form

\begin{eqnarray}
{\cal J}^\mu =\int dx^\mu (\tau)\, \delta^3[x-x(\tau )]\;,
\label{vort2}
\end{eqnarray}

\noindent
where $x^\mu(\tau )$ gives the vortex trajectory parametrized by $\tau$, $0
\leq \tau \leq 1$.  Using (\ref{vort2}), the action term in Eq. (\ref{S})
corresponding to the interaction of the vortex with the gauge field becomes

\begin{equation}
S_\mathrm{vortex,int} = i g \int \frac{d x^{\mu}}{d \tau} d \tau \, h_{\mu} \;,
\label{Svort}
\end{equation}
where $g = 2m/e$ is the vortex strength (again for unit winding number
vortices, which we are here considering).  Eq (\ref{Svort}) is the analogous
of a classical action of a charged particle, with charge $q\equiv g$ and null
rest mass, interacting with a vectorial field $h_{\mu}$.  Thus, technically,
we can identify the vortex as the worldline of a dual charged particle, which,
under second quantization, can be associated to a (local) complex vortex field
$\psi$ (this is much the same as used in many condensed matter applications of
duality for vortices, e.g. like in Refs.  \cite{dolan,balents}), where a
phenomenological field theory for vortices in two-dimensional space was
considered.  Here, in order to write a field model for vortices we follow the
approach adopted e.g. in Ref. \cite{kawai} for matching the action part
corresponding to the vortices degrees of freedom to a field theory model.
{}First, let us consider the exponential term in Eq. (\ref{finalZ}) given by
Eq. (\ref{S}) and write it in terms of the vacuum expectation value for the
scalar field, $v \equiv \rho_0/\sqrt{2}$,

\begin{eqnarray}
S&=& S_{\rm vortex,0} + \int d^3 x  \left[ \frac{m^2}{16 \pi^2 e^2
\rho_0^2}H_{\mu\nu}^2 -i \frac{m}{e} {\cal J}_\mu h_\mu +  i
\frac{m^2}{8 \pi^2 \theta} \epsilon_{\mu \nu \gamma} h_\mu
\partial_\nu h_\gamma
\right] \;,
\label{S2}
\end{eqnarray}
where

\begin{eqnarray}
S_{\rm vortex,0} & =& \int d^3 x  \left[ \frac{m^2}{16 \pi^2 e^2}
\left(\frac{1}{\rho^2} -\frac{1}{\rho_0^2} \right)
H_{\mu\nu}^2 + \frac{1}{2}\left( \partial_\mu \rho \right)^2 +
V(\rho)      \right]  \;.
\label{Sv0}
\end{eqnarray}
Note that in the approximation that the vortex is classical and fluctuations
of $\rho$ can be neglected (this is usually called in the condensed matter
problem as the London approximation \cite{kleinertI-II}), which should be
valid provided we remain deep inside the broken phase. This should be the case
for instance when thermal effects are not important. We next note that Eq.
(\ref{Sv0}) is only non-vanishing on the vortex core. Then, by properly
connecting Eq.  (\ref{Sv0}) in the dual transformed variables to its original
field form in terms of the vortex solutions, we can recognize that Eq.
(\ref{Sv0}) can be expressed back in the energy functional form for the
vortices and it can then be written in a Nambu like form
as~\cite{davis,vachaspati},

\begin{equation}
S_{\rm vortex,0} = E_{\rm vortex} \int d\tau\;,
\label{Sv02}
\end{equation}
where $E_{\rm vortex} = \pi \rho_0^2$, for unit-winding number self-dual
vortices.  Note that by associating Eq. (\ref{Sv0}) to the classical vortex
energy and the explicit use of the vortex solutions, a definite scale is been
introduced into the problem by the use of the characteristics of the classical
vortex solutions. It is natural and consistent, therefore, to consider from
this point onwards that we are effectively working with effective objects
described by thick vortices. The natural scale being introduced by the use of
the classical vortex solutions is the vortex radius, given in terms of the
Higgs mass in the broken phase as $a\sim 1/m_H$.

Considering the full vortex contribution to the partition function and the
action in the London approximation, then from Eq. (\ref{Sv02}) and the
interaction term with the dual gauge field $h_\mu$, Eq.  (\ref{Svort}), we
have that the total vortex contribution to the action is of the form

\begin{equation}
S_{\rm vortex}= E_{\rm vortex} \int d\tau +
i \frac{2 m}{e} \int \frac{d x^{\mu}}{d \tau} d\tau  h_{\mu}\;.
\label{Sv}
\end{equation}
This vortex action term can be matched to a field theory model for vortices,
described by a vortex field $\psi$ interacting with the gauge field $h_\mu$
and in such a way that the gauge symmetry of the action (\ref{S}) remains
preserved~\cite{kawai,seoSugamoto,rey} (see also e.g.  Ref. \cite{dual} for an
explicit derivation concerning vortex-strings excitations in 3+1 dimensions),

\begin{equation}
S_\mathrm{vortex} = \int d^3 x \left[ \left| \partial_{\mu}\psi + i\frac{2m}{e}
h_{\mu}\psi \right|^2 + M^2 |\psi|^2 \right] \;,
\label{Svort2}
\end{equation}
with the additional gauge invariance for the complex vortex field, $\psi(x)
\to \psi(x) \exp[- i 2m \Lambda(x)/e]$.  In the process of matching the
functional integration over the vortex coordinates in the original functional
partition function (\ref{finalZ}) to the one in terms of the vortex field, a
dynamical (entropy) mass term $M$ is induced. It is expressed in terms of the
classical vortex energy in Eq. (\ref{Sv}) and the characteristic length scale
for classical vortices, $a$, as~\cite{kawai}

\begin{equation}
M^2 = \frac{1}{a^2} \left(e^{a \, E_{\rm vortex}} -6\right)\;.
\label{M}
\end{equation}
As mentioned above, $a$ is set as the characteristic coherence length in the
vortex phase, which is given by the inverse of the scalar field Higgs mass in
the broken phase, $a\sim 1/m_H$, where $m_H$ is determined depending on the
scalar potential being used. {}For the self-dual potential Eq. (\ref{V2})
considered here, $m_H = e^2 \rho_0^2/\theta$.  Also, for self-dual vortices,
in Eq. (\ref{M}) we also have that $E_{\rm vortex} = \pi \rho_0^2$. Note that
by restricting the analysis to self-dual vortices, we do not need to consider
e.g. interacting terms for the vortices, that can be a complicate matter to
add to the field theory model Eq.  (\ref{Svort2}).

In principle, interaction terms for the vortices can be constructed in
general, for example by introducing in the vortex action a core energy term
for the vortices (see e.g. Refs.  \cite{kleinertI-II,kleinert} where this is
used) of the form $\sim \varepsilon {{\cal J}_\mu}^2$, which still preserves
the gauge symmetry for the total action term in the functional partition
function. The coupling $\varepsilon$ could be for instance phenomenologically
matched to some physical system of interest modeling two-dimensional systems
inspired by a CSH theory. In this work, however, we will not try to go that
far, but this could be an interesting possible extention of the approach used
here.

{}For the initial purposes set for this work, of confirming and determining
the location of the critical point for vortex condensation formally
demonstrated in Ref. \cite{caffarelli}, our result given by Eq. (\ref{M})
already suffices.  It follows from Eq.  (\ref{M}) that for model parameters
for which $M^2$ vanishes and then beyond that becomes negative, points to the
case where vortex excitations can condense, since a vortex condensate would be
energetically favorable to form.  {}For the self-dual potential considered, we
obtain that Eq. (\ref{M}) vanishes and change sign for CS parameters below a
critical value $\theta_c$ given by

\begin{equation}
\theta_c \simeq \frac{\ln 6}{\pi} e^2 \;.
\label{thetac}
\end{equation}
This result corroborates for instance the demonstration in Ref.
\cite{caffarelli} about the existence of a critical value for the CS parameter
below which vortex condensation should exist.  {}For potentials other than the
self-dual one, this critical value for the Chern-Simons coefficient can be a
complicate function of the model parameters, since we then need to know the
complete expression for the vortex free energy.  In the next section we
investigate whether quantum corrections to the dynamical vortex mass can
appreciably change the result given by Eq. (\ref{thetac}).

\section{The Vortex Field Self-Energy}
\label{self}

The analysis of the stability of the result (\ref{thetac}) towards quantum
corrections can be made by means of the evaluation of the self-energy quantum
contributions to the vortex field dynamical mass.  {}From the action
(\ref{S2}) and using the result (\ref{Svort2}), the action expressed in terms
of the dual gauge field $h_\mu$ and the vortex field $\psi$ becomes

\begin{eqnarray}
S&=& \int d^3 x  \left[ \frac{m^2}{16 \pi^2 e^2
\rho_0^2}H_{\mu\nu}^2
+  i \frac{m^2}{8 \pi^2 \theta} \epsilon_{\mu \nu \gamma} h_\mu
\partial_\nu h_\gamma \right. \nonumber \\
&+& \left. \left| \partial_{\mu}\psi + i\frac{2m}{e}
h_{\mu}\psi \right|^2 + M^2 |\psi|^2 + \frac{(\partial_\mu
h_\mu)^2}{2\alpha}\right] \;,
\label{S3}
\end{eqnarray}
where in the above equation $(\partial_\mu h_\mu)^2/(2\alpha)$ is a gauge
fixing term.

In the London approximation used to derive Eq. (\ref{S3}) the vortex field
only couples to the dual gauge field $h_\mu$. The interaction vertices
relevant for the calculation of the one-loop self-energy for the vortex field
$\psi$ come then from the terms $i(2 m/e) h_\mu \left[ \psi \partial_\mu
  \psi^* - \psi^* \partial_\mu \psi \right]$ and $(2 m/e)^2 h_\mu^2 |\psi|^2$.
The propagators for $\psi$ and $h_\mu$ can be determined from the quadratic
Lagrangian density in the form

\begin{equation}
\mathcal{L}_2= \frac{1}{2}h_{\mu}D_{\mu\nu}^{-1}h_{\nu} +
\psi D_{\psi\psi^*}^{-1} \psi^* \;,
\label{L2}
\end{equation}
where $D_{\mu\nu}^{-1}$, by redefining the $h_\mu$ field by a constant factor:
$h_\mu \to (2 \pi e \rho_0/m) h_\mu$ (note that this corresponds just to fix
$m^2$ to be $4\pi^2 e^2 \rho_0^2$), is given by

\begin{equation}
D_{\mu\nu}^{-1} = \left(1 -
\frac{1}{\alpha} \right)\partial_{\mu}\partial_{\nu} -
\partial^2\delta_{\mu\nu} - i\frac{e^2 \rho_0^2}{\theta}
\epsilon_{\mu \gamma \nu}\partial_{\gamma}\;,
\label{hh}
\end{equation}
while $D_{\psi\psi^*}^{-1}$ is given by

\begin{equation}
D_{\psi\psi^*}^{-1} = -\partial^2 + M^2\;.
\label{psipsi}
\end{equation}

The inverse of Eqs. (\ref{hh}) and (\ref{psipsi}) define the field propagators
for $h_\mu$ and $\psi$, respectively. In particular, for the $h_\mu$ field, we
have that the propagator, in momentum space, is given by

\begin{eqnarray}
D_{\mu\nu}(k) & = & \frac{\left[(\alpha - 1) k^2  + \alpha e^4
\rho_0^4/\theta^2\right] k_{\mu}k_{\nu} }{ k^4 \left( k^2 +
e^4 \rho_0^4/\theta^2 \right)} + \frac{
\left( e^2 \rho_0^2/ \theta\right)\epsilon_{\mu \nu \beta}
k_{\beta}}{k^2 \left( k^2 + e^4 \rho_0^4/\theta^2 \right)} \nonumber \\
& & + \frac{\delta_{\mu\nu}}{k^2 + e^4 \rho_0^4/\theta^2}\;,
\label{proph}
\end{eqnarray}
while for the vortex field it is just

\begin{equation}
D_{\psi\psi^*}(k) = \frac{1}{k^2 + M^2} \;.
\label{proppsi}
\end{equation}

A convenient choice of gauge in Eq. (\ref{proph}) is the Landau gauge $\alpha
= 0$. With this choice we have $k_{\nu}D_{\mu\nu} = 0$, which assures that all
contributions coming from derivative vertex Feynman diagrams vanish. In the
following we adopted the Laudau gauge in the calculation of the vortex
self-energy.

The diagrams contributing to the vortex self-energy, $\Sigma$, at one-loop
order are shown in {}Fig. 1.

\begin{figure}[htb]
  \vspace{0.5cm}
  \epsfig{figure=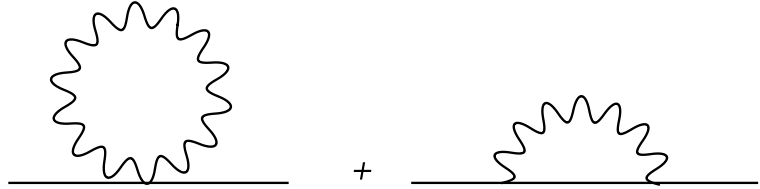,angle=0,width=10cm}
\caption[]{\label{fig1} Diagrams contributing to the vortex field
  self-energy at the one-loop order. The continuous line denotes the vortex
  field propagator, while the wavy line the dual gauge field propagator.}
\end{figure}

The self-energy at one-loop order, using the gauge field redefinition given
below Eq. (\ref{L2}) and the field propagators Eqs. (\ref{proph}) and
(\ref{proppsi}), is then given by

\begin{eqnarray}
\Sigma(p) &=& 16 \pi^2 \rho_0^2 \delta_{\mu \nu} \int_{\rm vort}
\frac{d^3 k}{(2 \pi)^3} D^{\mu \nu}(k)  - 16 \pi^2 \rho_0^2 \int_{\rm vort}
\frac{d^3 k}{(2 \pi)^3} (2 p+k)_\mu D_{\psi\psi^*}(p+k)
(2 p+k)_\nu D^{\mu \nu}(k) \;,
\label{Sigma}
\end{eqnarray}
where the index in the momentum integral, $\int_{\rm vort}$, is to remember
that the momentum integral is to be evaluated considering the case of
effective thick vortices, with characteristic scale set by the (inverse of
the) Higgs mass $m_H$. This then sets a momentum cutoff $\Lambda \equiv 1/a
=m_H$.

In terms of the self-energy $\Sigma(p)$, we define an effective dynamical mass
for the vortex as given by

\begin{equation}
M_{\rm eff}^2 = M^2 + \Sigma(M_{\rm eff})\;,
\label{gap}
\end{equation}
where the self-energy is to be evaluated on-shell. Equation (\ref{gap}) is a
gap equation that has to be evaluated for the effective mass $M_{\rm eff}$.
The critical point for vortex condensation is then defined by the value of the
Chern-Simons coefficient for which $M_{\rm eff}(\theta_c) =0$. The critical
point $\theta_c$ is then obtained from Eq. (\ref{gap}) as given by the
solution of

\begin{equation}
\left[M^2 + \Sigma(0)\right]\Bigr|_{\theta=\theta_c} =0\;.
\label{solvegap}
\end{equation}

By explicitly using the field propagators in Eq. (\ref{Sigma}) and solving the
gap equation at the critical point, we find that Eq. (\ref{solvegap}) is given
by the following simple equation to be solved for $\theta_c$,

\begin{equation}
M^2(\theta_c) + \frac{16 e^2 \rho_0^4}{\theta_c} \left(1-\frac{\pi}{4}
\right)=0 \;.
\label{eq1}
\end{equation}
Using the tree-level result for the dynamical vortex mass, Eq. (\ref{M}), we
can numerically solve Eq. (\ref{eq1}) for $\theta_c$ and obtain that

\begin{equation}
\theta_c \simeq 0.825 \, \frac{\ln 6}{\pi} e^2\;,
\end{equation}
which is about $17\%$ smaller than the tree-level result, Eq. (\ref{thetac}),
derived in the previous section. Higher loop terms to the self-energy should
lead to ${\cal O}(e^3)$ and higher corrections to this result and, thus, they
are not expected to change appreciably the leading order one-loop result for
$\theta_c$, at least for perturbatively small couplings.

\section{Conclusions}

In this work we have given the expression for the quantum partition function
for vortices in the context of the CSH model. This is realized by obtaining
the dualized version of the model, where the contribution of vortex
excitations are made apparent in the action and also their coupling with the
matter fields.  The procedure explained in Sec. III to obtain the dual action
allows us to take into account, in the functional path integral, the
contribution of not only constant vacuum field fluctuations but also those
nontrivial singular topological excitations that are known to exist in the
original model.

By restricting the study of the obtained dual action in the London limit for
the scalar Higgs field, $\rho = \rho_0 \equiv \langle \rho \rangle$ taken as
constant and considering the classical self-dual vortex solutions, the vortex
degrees of freedom in the partition function action can be matched to a field
theory model in terms of a vortex field with a dynamical mass for the
vortices. {}From the expression of this dynamical vortex mass, we have shown
that we can define and obtain the critical point for which vortices are
energetically favorable to condense. This determines a specific critical value
for the CS parameter, $\theta_c$. {}For values of $\theta < \theta_c$ vortex
condensation is favorable already at the tree level.  We have also considered
the one-loop corrections to the dynamical vortex mass term and derived the
self-energy contribution for the vortex field.  We have shown that the
critical point for vortex condensation slight decreases when the quantum
corrections are included, but the prediction for vortex condensation still
remains. Higher loop corrections are not expected to change in any appreciable
way our predictions and results, at least for perturbative values for the
gauge coupling constant.

It would be interesting to further investigate the vortex condensation problem
in the quantum theory by possibly including interaction terms for the vortices
({\it i.e.} using scalar potentials other than the self-dual one), in which
case the method developed here could be useful in modeling,
phenomenologically, two-dimensional systems based on the CSH model, like in
the study of condensed matter planar systems, which are of importance in the
study of high-temperature cuprate superconductors and in the fractional
quantum Hall effect.  The inclusion of finite temperature effects, and then
going beyond the London approximation considered in this work, would also be a
natural extension to be done.  We hope to pursue these and other problems
related to this work in the future.

\acknowledgments

R.O.R. is partially supported by Conselho Nacional de Desenvolvimento
Cient\'{\i}fico e Tecnol\'{o}gico (CNPq-Brazil) and {}Funda{\c {c}}{\~{a}}o de
Amparo {\`{a}} Pesquisa do Estado do Rio de Janeiro (FAPERJ).


\end{document}